\documentclass[twocolumn,floatfix,aps]{revtex4-2}
\usepackage{bm}% bold math
\usepackage{graphicx,color}
\usepackage{graphicx}% Include figure files
\usepackage{dcolumn}% Align table columns on decimal point
\usepackage{times}

\usepackage{amsmath,amssymb}

\begin{document}

\title{Mean field theory of chaotic insect swarms} %mean field scale free chaotic swarms %mean field theory  of chaotic insect swarms
\author{R. Gonz\'alez-Albaladejo}
\affiliation{Departamento de Matem\'atica Aplicada, Universidad Complutense de Madrid, 28040 Madrid, Spain}
\affiliation{Gregorio Mill\'an Institute for Fluid Dynamics, Nanoscience and Industrial Mathematics, Universidad Carlos III de Madrid, 28911 Legan\'{e}s, Spain}
\author{L. L. Bonilla$^*$}
\affiliation{Gregorio Mill\'an Institute for Fluid Dynamics, Nanoscience and Industrial Mathematics, Universidad Carlos III de Madrid, 28911 Legan\'{e}s, Spain}
\affiliation{Department of Mathematics, Universidad Carlos III de Madrid, 28911 Legan\'{e}s, Spain. 
$^*$Corresponding author. E-mail: bonilla@ing.uc3m.es}
\date{\today}
\begin{abstract}
The harmonically confined Vicsek model displays qualitative and quantitative features observed in natural insect swarms. It exhibits a scale free transition between single and multicluster chaotic phases. Finite size scaling indicates that this unusual phase transition occurs at zero confinement [Physical Review E {\bf 107}, 014209 (2023)]. While the evidence of the scale-free-chaos phase transition comes from numerical simulations, here we present its mean field theory. Analytically determined critical exponents are those of the Landau theory of equilibrium phase transitions plus dynamical critical exponent $z=1$ and a new critical exponent $\varphi=0.5$ for the largest Lyapunov exponent. The phase transition occurs at zero confinement and noise in the mean field theory. The noise line of zero largest Lyapunov exponents informs observed behavior: (i) the qualitative shape of the swarm (on average, the center of mass rotates slowly at the rate marked by the winding number and its trajectory fills compactly the space, similarly to the observed condensed nucleus surrounded by vapor), and (ii) the critical exponents resemble those observed in natural swarms. Our predictions include power laws for the frequency of the maximal spectral amplitude and the winding number.

\end{abstract}

\maketitle

Collective animal motion has common features that suggest underlying principles beyond biological details \cite{oku86,par99,sum10,vic12, oue22}. Indeed, insect swarms, fish schools, bird and sheep flocks, or crowds of people exhibit collective properties distinct from those of their component individuals. In the past, these properties have been characterized as of hypnotic \cite{lebon1895} or telepathic \cite{long1919} nature, although other authors in those years  argued that the spread of impulse in well organized groups was adequate to explain the existence of a collective mind of the flock \cite{mil1921}. More recently, advances in stereo videography and calibration \cite{the14} have generated enormous amounts of quantitative data for collective animal motion \cite{oue22}. In particular, the observation of power laws and critical exponents in biological systems \cite{mor11,bia12,tan17,sum10,aza18,cav18} has generated much theoretical investigation into the unusual phase transitions which may be responsible for them. Power laws for natural insect swarms are deduced from correlation functions \cite{att14plos,att14,cav17,cav18}. Their relation to possible renormalization group theories of phase transitions \cite{wil83} have led to efforts to identify their postulated universality class \cite{cav17,che15,che18,cav21prr,cav21arxiv}. 

Besides observations in natural settings, mating swarms of male midges have been much studied in laboratory conditions \cite{oue22,kel13,puc15,ni15,gor16,sin17,fen23}. Midges perceive acoustic signals and move with low frequency maneuvers but react with synchronized high frequency oscillatory motion to the presence of nearby insects \cite{puc15}. When undriven, the swarm center of mass moves {\em almost randomly} on a plane (with larger fluctuations in the vertical direction of gravity) but it follows an elliptic trajectory when driven at 1 Hz frequency superimposed to the sound of a male midge \cite{ni15}. Motions of single midges in swarms follow L\'evy walks \cite{rey16}, which might indicate chaotic motion in related animal patterns \cite{rey16b}. Acoustic interaction of midges has been modeled by adaptive gravity, which produces an effective harmonic potential near the swarm center \cite{gor16}. Swarms comprise a core condensed phase surrounded by a dilute vapor phase with midges entering and leaving the core \cite{sin17}, while individual midges do not sample the swarm uniformly \cite{fen23}. The long range correlations between midges in the wild \cite{att14plos,att14,cav17,cav18} are not observed in laboratory conditions where background noise and atmospheric conditions are absent \cite{ni15epj}.  

Finding models accounting for all the different swarm features is challenging. Recently, we have discovered a phase transition in the harmonically confined Vicsek model (HCVM), characterized by scale free chaos, which exhibits several observed traits of the swarm (condensed nucleus and vapor phases, flatness at the origin and,  on a bounded interval, collapse of dynamic correlation function in terms of time divided by correlation length) and is compatible with observed critical exponents \cite{gon23}. For finitely many insects, the scale-free-chaos phase transition is a critical line separating single from multicluster chaotic swarms, and having correlation length proportional to swarm size. This line converges to zero confinement as insect number goes to infinity and chaos disappears \cite{gon23}. Since our findings are based on numerical simulations, it is important to have a theory to interpret them. Not having a renormalization group theory of the HCVM scale-free-chaos phase transition, we develop here a mean field theory of the HCVM as a first step. Note that the standard Vicsek model with periodic boundary conditions \cite{vic95,vic12} displays an ordering transition \cite{cha20} that is very different from the HCVM scale-free-chaos transition \cite{gon23}.

The three dimensional (3D) HCVM satisfies
\begin{eqnarray}
&&\mathbf{x}_i(t+1)=\mathbf{x}_i(t)+ \mathbf{v}_i(t+1),\quad i=1,\ldots,N,\nonumber\\
&& \mathbf{v}_i(t+1)=v_0  \mathcal{R}_\eta\!\left[\Theta\!\left(\sum_{|\mathbf{x}_j-\mathbf{x}_i|<R_0}\mathbf{v}_j(t)-\beta\mathbf{x}_i(t)\right)\!\right]\!, \label{eq1}
\end{eqnarray}
where $\Theta(\mathbf{x})=\mathbf{x}/|\mathbf{x}|$, $R_0$ is the radius of the sphere of influence about particles, $\beta$ is the confining spring constant, and $\mathcal{R}_\eta(\mathbf{w})$ performs a random rotation uniformly distributed around $\mathbf{w}$ with maximum amplitude of $\eta$ \cite{gon23}. Firstly, we set $\eta=0$, average these equations using the definition
\begin{eqnarray}
\langle f(\mathbf{x}_i)\rangle = \frac{1}{N}\sum_{i=1}^N f(\mathbf{x}_i),\quad \mathbf{X}(t)=\langle\mathbf{x}_i\rangle,   \label{eq2}
\end{eqnarray}
in the limit as $N\to\infty$, and assume the mean field (tree \cite{ami05}) approximation $\langle f(\mathbf{x}_i)\rangle \approx f(\langle\mathbf{x}_i\rangle)$. The result is
\begin{eqnarray}
&&\mathbf{X}(t+1)-\mathbf{X}(t)= v_0\Theta\!\left( \mathbf{X}(t)-\mathbf{X}(t-1)-\tilde{\beta}\mathbf{X}(t)\right)\!,\quad \label{eq3}
\end{eqnarray}
where $\tilde{\beta}=\beta/M$ and $M$  is the average number of particles within the sphere of influence about $i$, all of which remain inside the sphere. We have used that, for a compact swarm, $\left\langle\sum_{|\mathbf{x}_j-\mathbf{x}_i|<R_0}\mathbf{v}_j(t)\right\rangle=\left\langle\sum_{|\mathbf{x}_j-\mathbf{x}_i|<R_0}[\mathbf{x}_j(t)-\mathbf{x}_j(t-1)]\right\rangle\approx M[\langle\mathbf{x}_i(t)\rangle-\langle\mathbf{x}_i(t-1)\rangle]= M [\mathbf{X}(t)-\mathbf{X}(t-1)]$. Moreover, the initial positions $\mathbf{X}(0)$ and $\mathbf{X}(1)$ characterize a plane to which all successive positions given by \eqref{eq3} belong. This is similar to observed swarm motion \cite{ni15}. Restoring the alignment noise in \eqref{eq3}, we obtain the {\em stochastic mean field HCVM} (MFHCVM):
\begin{subequations}\label{eq4}
\begin{eqnarray}
&&\mathbf{X}(t+1)=\mathbf{X}(t)+ \mathbf{V}(t+1), \label{eq4a}\\
&& \mathbf{V}(t+1)= v_0\mathcal{R}_\eta\!\left[ \Theta\!\left( \mathbf{V}(t)-\tilde{\beta}\,\mathbf{X}(t)\right)\right]\! . \label{eq4b}
\end{eqnarray}
\end{subequations}

\begin{figure}[h]
\begin{center}
\includegraphics[width=7cm]{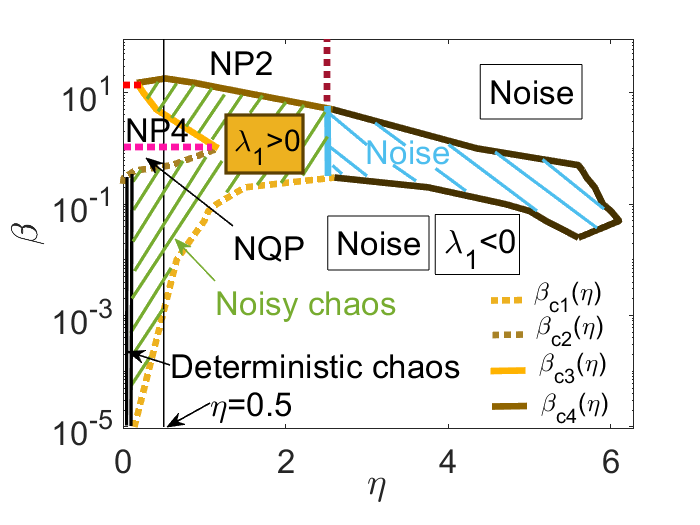}
\end{center}
\caption{Phase diagram on confinement vs noise plane indicating regions of deterministic and noisy chaos, noisy period-$\sigma$ (NP$\sigma$) and noisy quasiperiodic (NPQ) attractors, and mostly noise. We set $M=v_0=1$, effectively replacing $\beta$ instead of $\tilde{\beta}$ in all figures. \label{fig1}}
\end{figure}
Numerical simulations of the MFHCVM produce the phase diagram in Fig.~\ref{fig1}. For $\eta=0$, deterministic chaos with positive largest Lyapunov exponent (LLE) $\lambda_1$ occurs on the interval $(0,\beta_{c2})$. For nonzero noise, noisy chaos (as defined from scale-dependent Lyapunov exponents \cite{gao06,gon23}) appears on the intervals $(\beta_{c1}(\eta),\beta_{c2}(\eta))$ and $(\beta_{c3}(\eta),\beta_{c4}(\eta))$, whereas noisy quasiperiodic and periodic attractors exist elsewhere \cite{suppl}. At $\beta=\eta=0$, $\lambda_1 = 0$. For $2<\eta<2\pi$, noise dominates even though there is a region of chaos swamped by noise for intermediate values of $\beta$. For the MFHCVM, the scale-free-chaos phase transition of the 3D HCVM \cite{gon23} corresponds to the origin in Fig.~\ref{fig1}. 

\begin{figure}[h]
\begin{center}
\includegraphics[width=7cm]{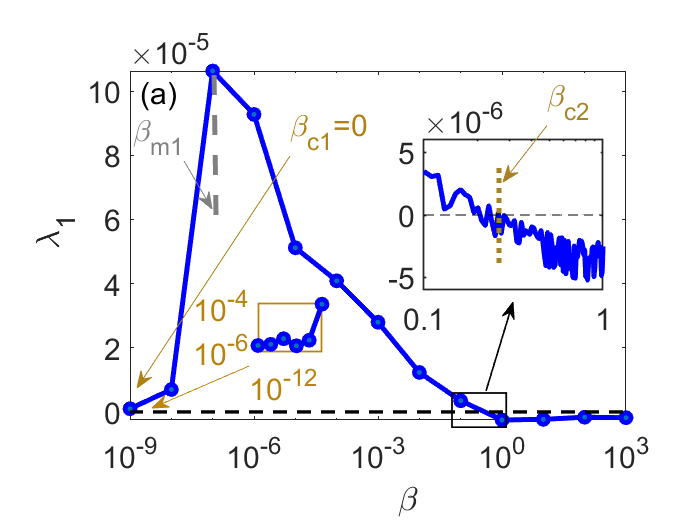}
\includegraphics[width=7cm]{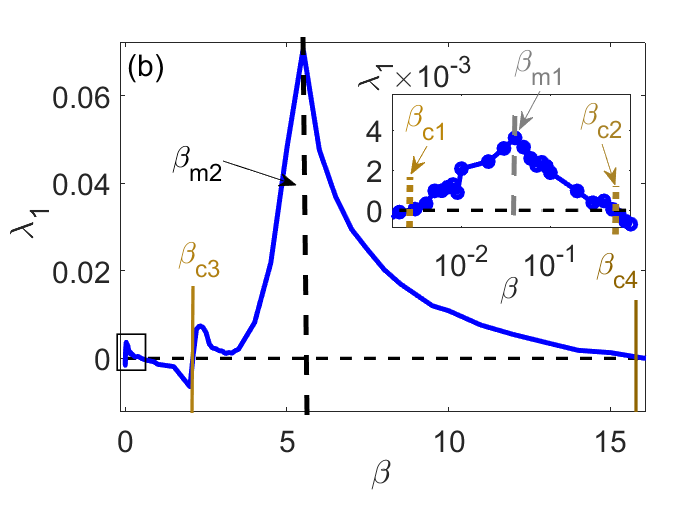}
\end{center}
\caption{Largest Lyapunov exponent versus confinement for {\bf (a)} $\eta=0$, {\bf (b)} $\eta=0.5$ showing the windows of deterministic and noisy chaos, respectively. The insets show zooms of the rectangular regions in the main figures of each panel. \label{fig2}}
\end{figure}

Figure \ref{fig2} shows the windows of positive LLE for vertical lines in Fig.~\ref{fig1} at noises $\eta=0$ and $\eta=0.5$, corresponding to deterministic and noisy chaos, respectively. For zero noise, the chaotic window ends at $\beta_{c2}=0.2$ and begins at $\beta_{c1}=0$ (in our simulations, we still get a clearly positive LLE at $\beta= 10^{-9}$). There are two chaotic windows $(\beta_{c1},\beta_{c2})$ and $(\beta_{c3},\beta_{c4})$ for $\eta=0.5$ in Fig.~\ref{fig2}(b). Inside these windows, the LLE has  peaks at $\beta_{m1}$ and $\beta_{m2}$, respectively. Fig.~\ref{fig1} shows that the scale-free-chaos phase transition is located at the origin of the phase diagram $(\beta,\eta)$. While we can reach this transition by lowering $\beta$ at $\eta=0$, we can also move on the critical line $\beta_{c1}(\eta)$ in Fig.~\ref{fig1} and let $\eta\to 0$ until we reach the origin. This latter route to the scale-free-chaos phase transition is reminiscent of finite size scaling for the HCVM, which produces critical exponents by letting $N\to\infty$ on the critical lines $\beta_c(N;\eta)$ as $\beta_c\to 0+$ having fixed $\eta$ on the region of noisy chaos \cite{gon23}. The critical exponents obtained by either route are the same but the deterministic route is more amenable to theory, whereas the critical exponents obtained by descending through the noise line $\beta_{c1}(\eta)$ in Fig.~\ref{fig1} follow from numerical simulations.

\begin{widetext}
The maximum at $\beta_{m1}\approx 10^{-7}$ in Fig.~\ref{fig2}(a) is in a region of chaotic attractors filling a large portion of space (for $\beta<10^{-6}$). The chaotic attractors fill a smaller annular region for $\beta>10^{-6}$; see Fig.~\ref{fig3}(a). Attractors filling large regions of space have zero average position and velocity but nonzero time averaged amplitudes $\langle|\mathbf{X}(t)|^2\rangle_t$, $\langle|\mathbf{V}(t)|^2\rangle_t=v_0^2$.
\begin{figure}[h]
\begin{center}
\includegraphics[width=10.5cm]{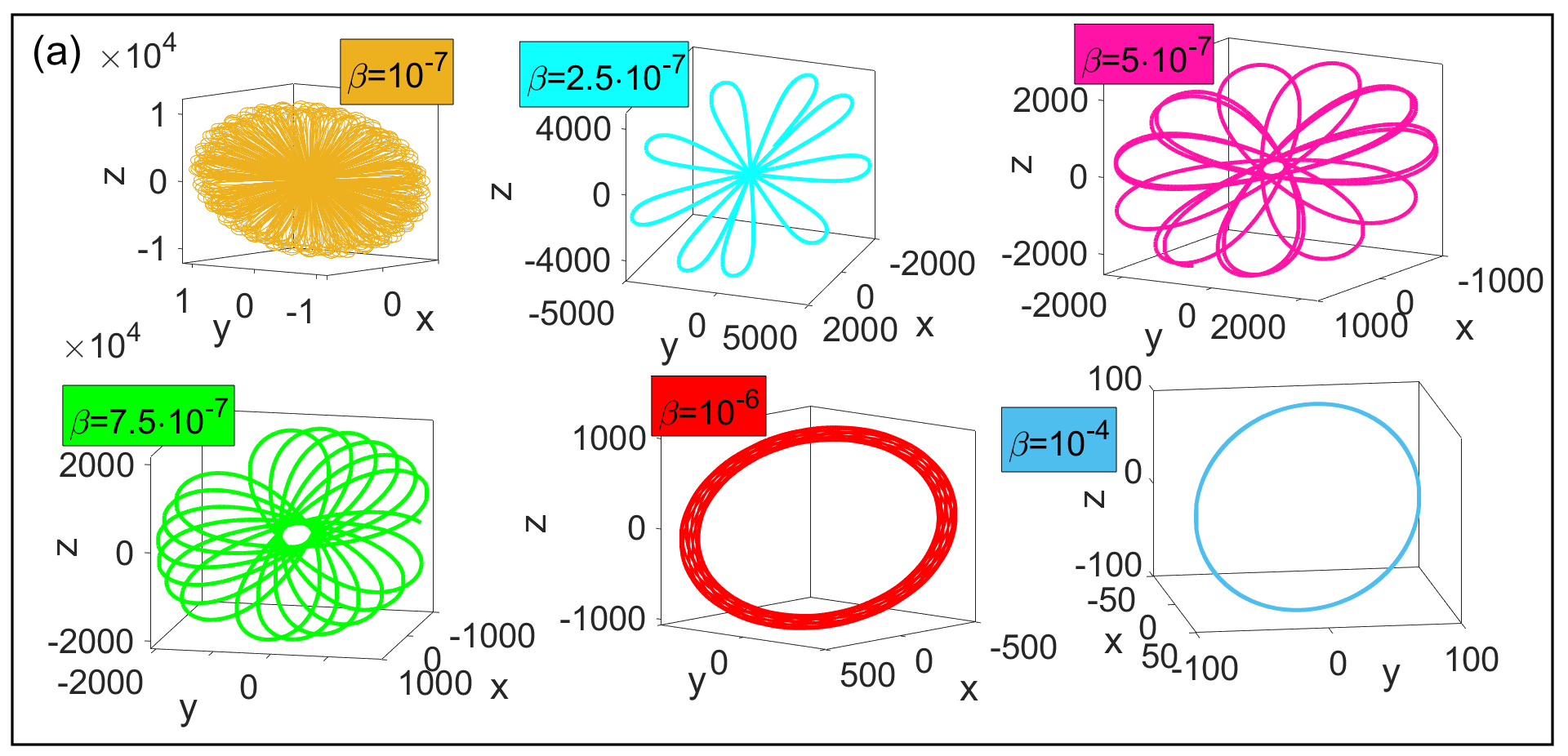}
\includegraphics[width=7cm]{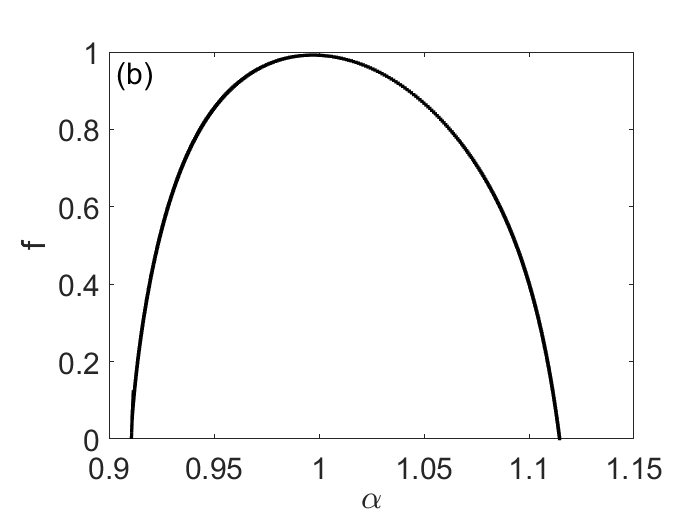}
\end{center}
\caption{Deterministic dynamics. {\bf (a)} Chaotic attractors (depicted on a short time interval) for different values of $\beta$. {\bf (b)} Multifractal singularity spectrum for the transition between quasiperiodicity and chaos. \label{fig3}}
\end{figure}
\end{widetext}
Fig.~\ref{fig3}(b) shows the singularity spectrum $f(\alpha)$ \cite{hal86,ott93,cen10} of the transition from quasiperiodicity to chaos at $\beta_{c2}(0)$  of Fig.~\ref{fig2}(a). Its shape is that of the circle map, with a maximum $D_0=\max_\alpha f(\alpha) = 1$ equating the Haussdorff fractal dimension at the transition. The zeros $D_\infty<D_{-\infty}$ of $f(\alpha)$ do not take on the same values as those for the golden-mean winding number of the critical circle map, $D_{\infty}\approx 0.6326$ and $D_{-\infty}\approx 1.8980$ \cite{hal86}: $(D_\infty,D_{-\infty})$ is narrower for the undriven transition between quasiperiodicity and chaos at $\beta_{c2}(0)$.

We now find the critical exponents at $\beta=0$, which correspond to the scale-free-chaos phase transition of the HCVM \cite{gon23}. In the latter, the correlation length $\xi$ is proportional to swarm size, which can be defined as the time average of $R(t)=|\mathbf{X}(t)|$ \cite{suppl}. From \eqref{eq4a}, we obtain
\begin{eqnarray*}
v_0^2\!=|\mathbf{X}(t)|^2\! +|\mathbf{X}(t+1)|^2\! - 2|\mathbf{X}(t)|\,|\mathbf{X}(t+1)|\cos\theta(t+1)\\
=R(t)^2+R(t+1)^2-2R(t)R(t+1)\cos\theta(t+1),
\end{eqnarray*}
where $\theta(t+1)$ is the angle between $\mathbf{X}(t)$ and $\mathbf{X}(t+1)$. Averaging over time and ignoring fluctuations,
\begin{eqnarray*}
\langle R(t)^2\rangle_t\!&\approx&\!\langle R(t)\rangle_t^2, \langle R(t)R(t+1)\cos\theta(t+1)\rangle_t\\
\!&\approx&\! \langle R(t)\rangle_t^2\cos\langle\theta(t)\rangle_t=\langle R(t)\rangle_t^2\cos(2\pi w),
\end{eqnarray*}
where $w$ is the winding number \cite{suppl}. Thus, we get $v_0^2\approx 2\langle R\rangle_t^2[1-\cos(2\pi w)]$. As $\beta\to 0+$, $w\to 0$ (see Fig.~\ref{fig4}), and therefore $\langle R(t)\rangle_t\to\infty$ with 
\begin{eqnarray}
w\sim \frac{v_0}{2\pi \langle R\rangle_t}. \label{eq5}
\end{eqnarray}
For zero noise, winding number and the frequency of the highest peak of the power spectrum, $\Omega=w$, for a signal $s(t)=X(t)+Y(t)+Z(t)$ coincide; see Fig.~\ref{fig4}(a). $\Omega$ is the reciprocal correlation time, therefore \eqref{eq5} implies $\tau\sim\xi$ (relating correlation time and length), and the dynamical critical exponent is $z=1$. For nonzero noise, the relation between winding number and peak frequency $\Omega$ is more complex; see Fig.~\ref{fig4}(b). Within the first chaotic window, $(\beta_{c1},\beta_{c2})$, $\Omega=w$, whereas $\Omega<w$ for $\beta<\beta_{c1}$. Within the second chaotic window, $(\beta_{c3},\beta_{c4})$, $\Omega$ is piecewise constant, with a finite jump at $\beta_{m2}$, corresponding to the maximum of the LLE. The winding number is smooth: it is slightly larger than $\Omega$ for  $\beta_{c3}<\eta<\beta_{m2})$, and it is slightly smaller than $\Omega$ for  $\beta_{m2}<\beta< \beta_{c4})$; see the inset of Fig.~\ref{fig4}(b).

To obtain the other critical exponents, we note that the order parameter of the MFHCVM cannot be the polarization because $|\mathbf{V}(t)|/v_0=1$. For $\eta=0$, \eqref{eq4b} yields
\begin{widetext}
\begin{eqnarray*}
R(t+1)^2= R(t)^2+v_0^2+2v_0\frac{(1-\tilde{\beta})R(t)^2-R(t)R(t-1)\cos\theta(t)}{\sqrt{(1-\tilde{\beta})^2R(t)^2+R(t-1)^2 -2R(t)R(t-1) (1-\tilde{\beta})\cos\theta(t)}}.
\end{eqnarray*}
\begin{figure}[h]
\begin{center}
\includegraphics[width=7cm]{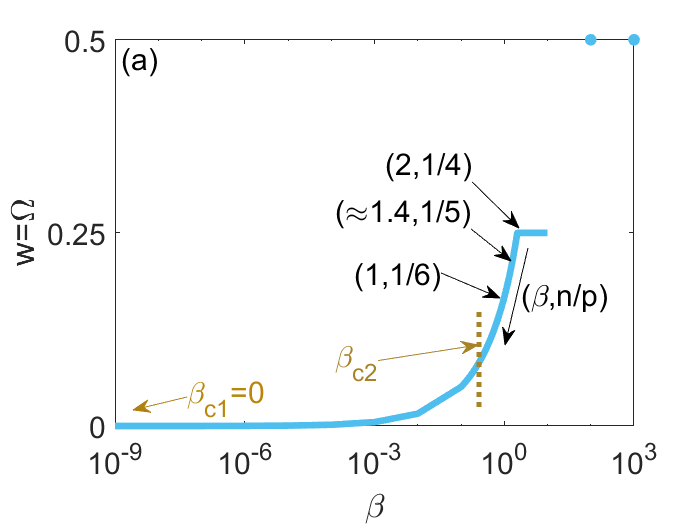}
\includegraphics[width=7cm]{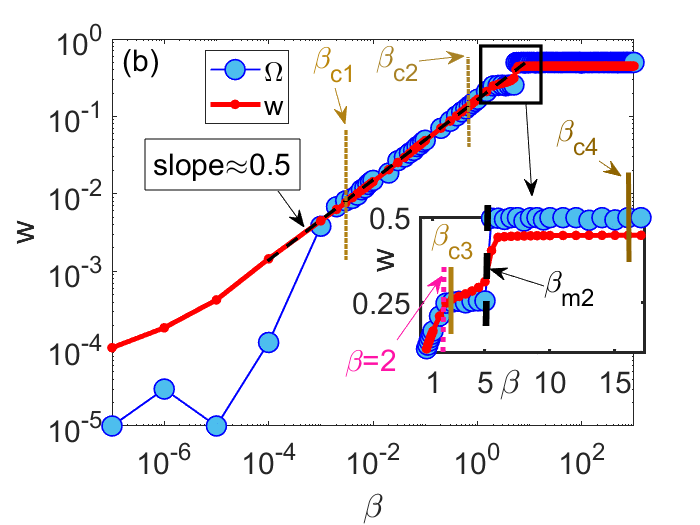}
\end{center}
\caption{Winding number versus confinement for {\bf (a)} $\eta=0$, {\bf (b)} $\eta=0.5$ indicating periodic, quasiperiodic attractors and ends of the chaotic windows as $\beta$ decreases. \label{fig4}}
\end{figure}
\end{widetext}
Time averaging this expression and ignoring fluctuations, 
\begin{eqnarray*}
0=v_0^2+2v_0\langle R\rangle_t\frac{1-\tilde{\beta} -\cos(2\pi w)}{\sqrt{2(1-\tilde{\beta})[1-\cos(2\pi w)]+\tilde{\beta}^2}}.
\end{eqnarray*}
In the limit as $\beta\to 0+$, $\langle R\rangle_t\to\infty$, $w\to 0$, and this equation yields $1 -\cos(2\pi w)\sim\tilde{\beta}-v_0\sqrt{2[1-\cos(2\pi w)]}/(2 \langle R \rangle_t)$, from which
\begin{eqnarray}
w\sim\sqrt{\frac{\tilde{\beta}}{4\pi^2}}. \label{eq6}
\end{eqnarray}
Thus, we have found the relation $w\sim\beta^b$, with $b=1/2$, for the winding number, which plays the role of order parameter in the Landau theory. From \eqref{eq5} and \eqref{eq6}, we find the critical exponent $\nu=1/2$ in the relation $\xi\sim\tilde{\beta}^{-\nu}$. We define the susceptibility $\chi$ as the time averaged norm of the linear response matrix $\mathbf{\mathcal{H}}_t=\nabla_\mathbf{H}\bm{\chi}_t= \partial\bm{\chi}_t^i/\partial H_j$ (at zero field) to an external force resulting from replacing $\mathbf{V}(t)+\mathbf{H}$ instead of the alignment force $\mathbf{V}(t)$ in \eqref{eq4b}\cite{suppl}:
\begin{eqnarray}
\chi=\langle\lVert\mathbf{\mathcal{H}}_t\rVert\rangle_t, \quad \lVert\mathbf{\mathcal{H}}_t\rVert =\sqrt{\lambda_M(\mathbf{\mathcal{H}}_t\mathbf{\mathcal{H}}_t^T)}. \label{eq7}
\end{eqnarray}
Here $\lambda_M(\mathbf{\mathcal{H}}_t\mathbf{\mathcal{H}}_t^T)$ is the maximum eigenvalue of the symmetric positive matrix $\mathbf{\mathcal{H}}_t\mathbf{\mathcal{H}}_t^T$. Eq.~\eqref{eq4a} produces $\mathbb{Y}_{t+1}= \mathbb{Y}_t+ \mathbb{W}_{t+1}$, where $\mathbb{Y}^{ij}=(\mathbf{Y}^j )_i= \partial\mathbf{X}_i/\partial H_j$, and $\mathbb{W}^{ij}= (\mathbf{W}^j )_i=\partial\mathbf{V}_i/\partial H_j$, at $\mathbf{H}=\mathbf{0}$. Eq.~\eqref{eq4b} yields \cite{suppl}
\begin{widetext}
\begin{subequations}\label{eq8}
\begin{eqnarray} 
&&(\mathbf{W}_{t+1}^j)_i=\mathcal{R}_\eta\!\left(\mathbb{A}_t^{ik}\!\left[(\mathbf{W}^j_t)_k -\tilde{\beta}(\mathbf{Y}_t^j)_k\right]\! + \mathbb{A}_t^{ik}(\bm{\delta}^j)_k\right)\!,\quad (\bm{\delta}^j)_i= \delta_{ij},  \label{eq8a}\\
&&\mathbb{A}_t = \frac{v_0}{\left|\mathbf{V}(t)-\tilde{\beta}\mathbf{X}(t)\right|}\left[\mathbb{I}-\frac{\left(\mathbf{V}(t)-\tilde{\beta}\mathbf{X}(t)\right)\left(\mathbf{V}(t)-\tilde{\beta}\mathbf{X}(t)\right)^T}{\left|\mathbf{V}(t)-\tilde{\beta}\mathbf{X}(t)\right|^2}\right]\!. \label{eq8b}\end{eqnarray}\end{subequations}\end{widetext}
Here sum over repeated indices is understood. A modification of the argument leading to \eqref{eq6} produces $\chi\sim\frac{\partial \langle R\rangle_t}{\partial H}|_{H=0}\sim\tilde{\beta}^{-\gamma}$ with critical exponent $\gamma=1$ \cite{suppl}. Hence, the critical exponents of the MFHCVM are the same as those in the Landau theory of phase transitions \cite{hua87}:
\begin{equation}
\nu=b=0.5, \quad \gamma=1, \label{eq9}
\end{equation}
with dynamical critical exponent $z=1$. Fig.~\ref{fig5} exhibits power laws with critical exponents \eqref{eq9} and $z=1$ as obtained from numerical simulations. As we approach the origin in Fig.~\ref{fig1} through the line $\beta_{c1}(\eta)$ ($\eta\to 0$), we obtain the same critical exponents from numerical simulations, as shown in Fig.~\ref{fig6}. 

\begin{widetext}
For the deterministic and stochastic cases, the swarm size $L=\max_tR(t)$ is proportional to the time averaged length of the center of mass position as the confinement decreases; see Fig.~\ref{fig5}(a). Thus, swarm size and correlation length are proportional. Figs.~\ref{fig6}(a) shows the power law of the winding number, $w_{c1}\sim\beta_{c1}^b$, $b\approx 0.5$. Figs.~\ref{fig5}(b) and \ref{fig6}(b) plot the correlation length power law with critical exponent $\nu=0.5$. The dynamical critical exponent $z$ is found from the relation $\Omega\sim\xi^{-z}$, $z=1.01\pm 0.01$, between the frequency corresponding to the maximum of the power spectrum and the correlation length; see Figs.~\ref{fig5}(c) and \ref{fig6}(c). Lastly the power laws for the susceptibility are shown in Figs.~\ref{fig5}(d) and \ref{fig6}(d).

%Figure \ref{fig2} depicts the power laws for the stochastic case as we approach the origin of the phase diagram (Figure 1 of the main text) through the line $\beta_{c1}(\eta)$ that marks the beginning of the first chaotic window. We check that the critical exponents of Eq.~\eqref{eq25} are also obtained as $\beta\to 0$ on the line $\beta_{c1}(\eta)$. Since this line marks the beginning of the first chaotic window, the LLE is zero on it, and we cannot determine the critical exponent $\varphi$ for nonzero noise from it.
\begin{figure}[h]
\begin{center}
\includegraphics[width=4.4cm]{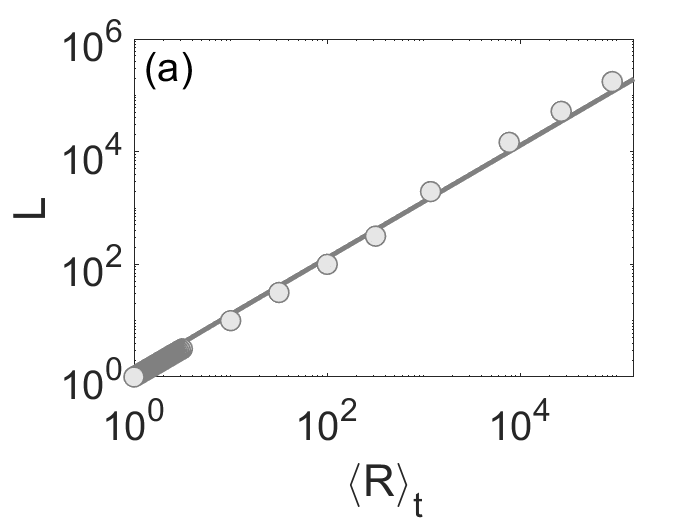}
\includegraphics[width=4.4cm]{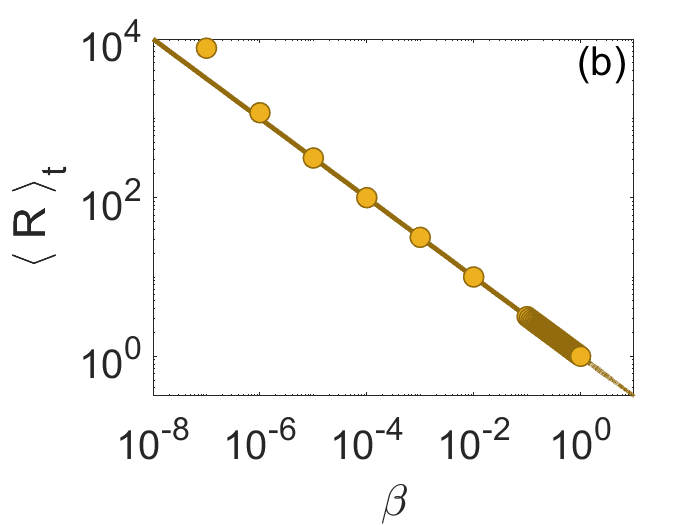}
\includegraphics[width=4.4cm]{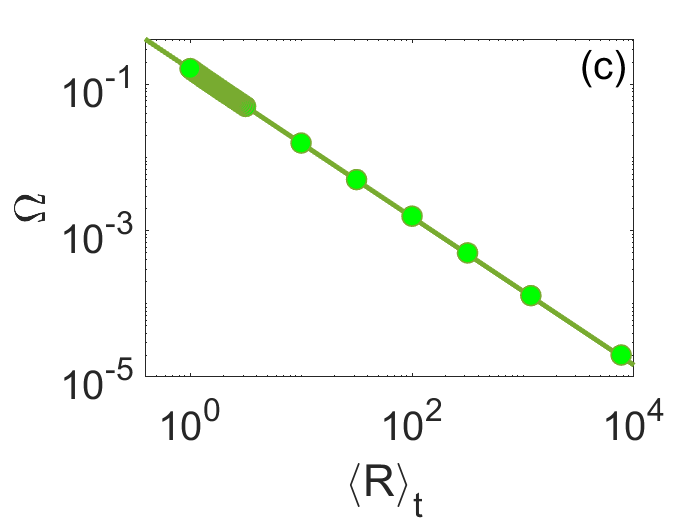}
\includegraphics[width=4.4cm]{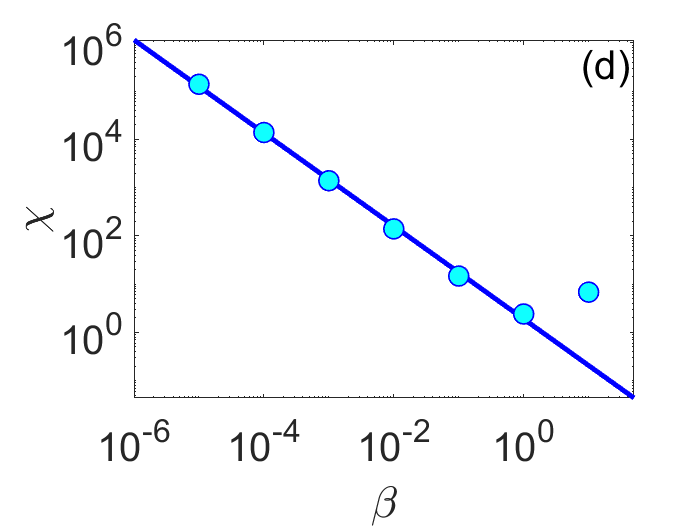}
\end{center}
\caption{Power laws for the deterministic case $\eta=0$ as we approach zero confinement, which corresponds to the scale-free-chaos transition of the harmonically confined Vicsek model. {\bf (a)} The maximum swarm size, $L=\mbox{max}(R)$, and the correlation length, $\xi=\langle R\rangle_t$ are proportional. {\bf (b)} The correlation length scales as $\xi\sim\beta^{-\nu}$ with $\nu=0.5$. {\bf (c)} The relation between the frequency for the maximum of the power spectrum scales as $\Omega\sim L^{-z}$ with the dynamic critical exponent $z=1.01\pm0.01$; and {\bf (d)} the susceptibility scales as $\chi\sim\beta^{-\gamma}$ with $\gamma=1.000\pm0.002$.} \label{fig5}
\end{figure}

\begin{figure}[h]
\begin{center}
\includegraphics[width=4.4cm]{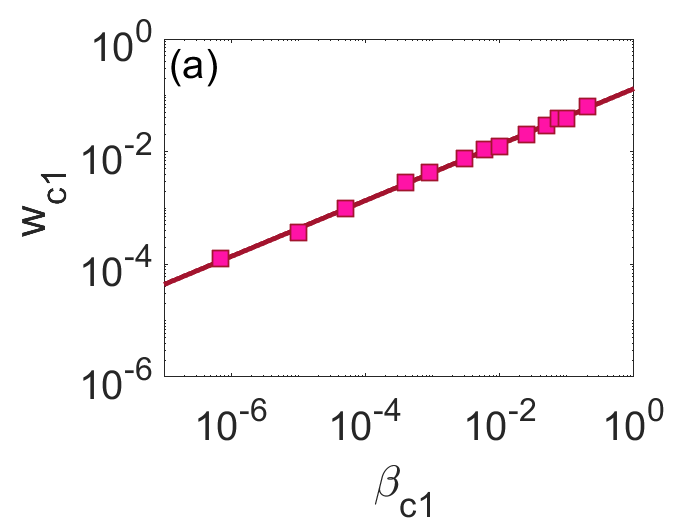}
\includegraphics[width=4.4cm]{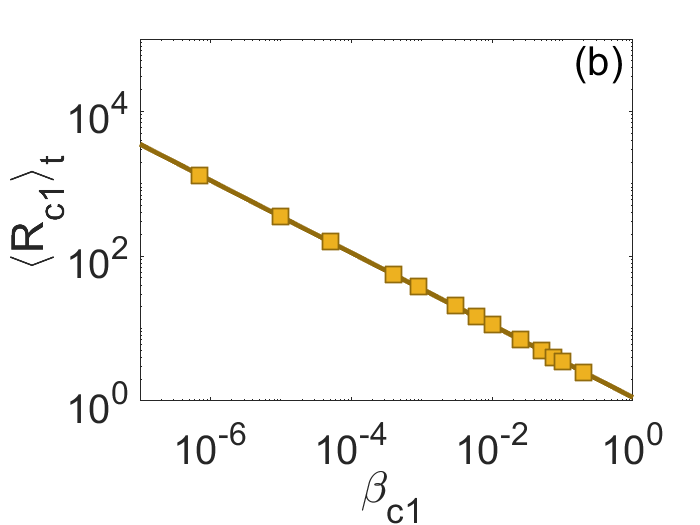}
\includegraphics[width=4.4cm]{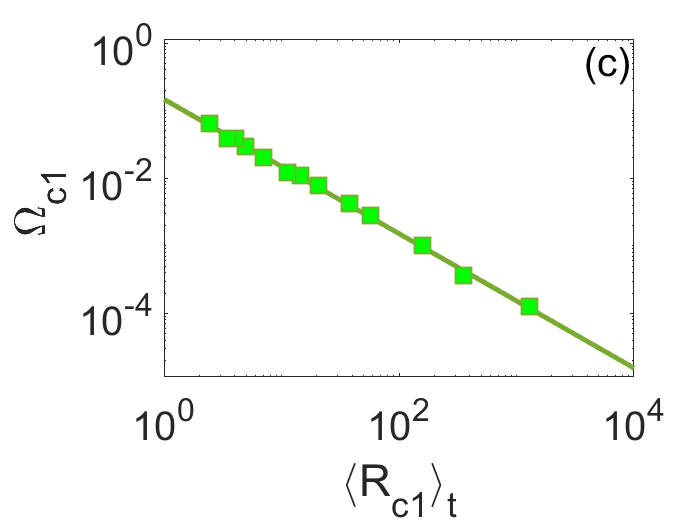}
\includegraphics[width=4.4cm]{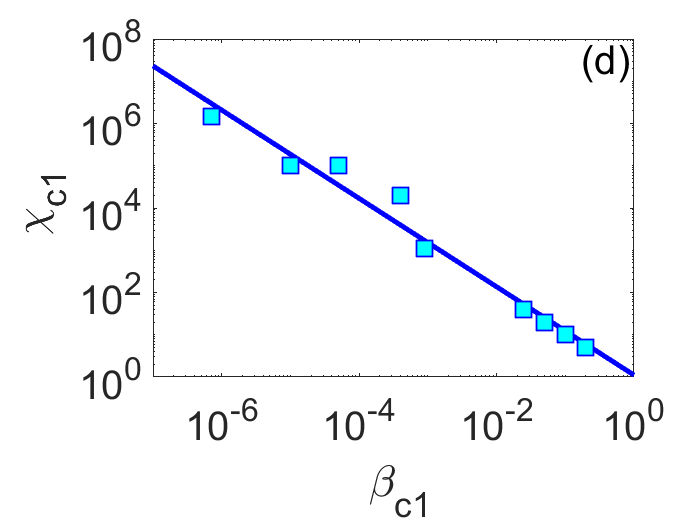}
\end{center}
\caption{Same as Figure \ref{fig5} for the stochastic case with $\beta=\beta_{c1}(\eta)$ and $\eta\to 0$. {\bf (a)} Power law for the order parameter, which is the winding number, versus $\beta$, $w\sim\beta^b$, with $b=0.497\pm 0.006$; {\bf (b)} Correlation length $\xi\sim\beta^{-\nu}$ with $\nu=0.500\pm 0.001$; {\bf (c)} dynamic critical exponent $z=0.99\pm 0.01$ for the law $\Omega\sim \xi^{-z}$; {\bf (d)} susceptibility vs confinement $\chi\sim\beta^{-\gamma}$ with $\gamma=1.04\pm 0.06$.} \label{fig6}
\end{figure}
\end{widetext}

\begin{figure}[h]
\begin{center}
\includegraphics[width=7cm]{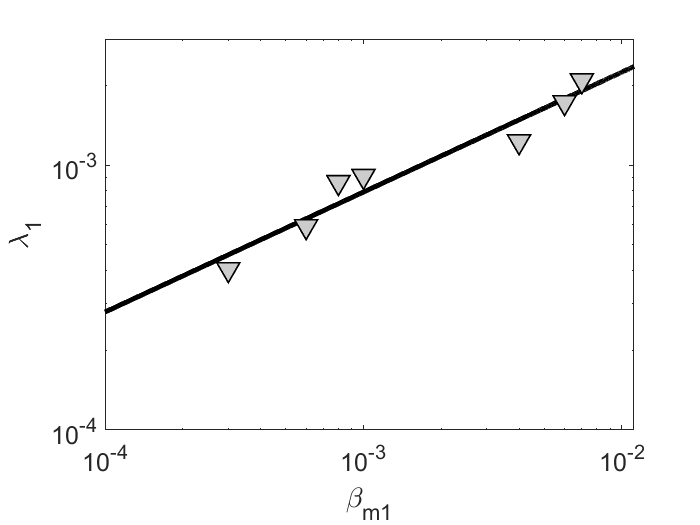}
\end{center}
\caption{LLE vs $\beta_{m1}(\eta)$, the location of the local maximum of the LLE at each value of the noise (stochastic MFHCVM) as $\eta\to 0$. The power law is $\lambda_1\sim\beta^\varphi$ with critical exponent $\varphi=0.45\pm 0.05$, which is compatible with the predicted value 0.5.} \label{fig7}
\end{figure}

In \cite{suppl}, we find the bound $\lambda_1\geq\sqrt{\tilde{\beta}}$ for the LLE. Assuming LLE satisfies a power law $\lambda_1\sim\tilde{\beta}^\varphi$, we obtain $\varphi\leq 0.5$. Finite velocity propagation implies $\varphi\geq\nu$ \cite{gon23}, and therefore $\varphi=\nu=0.5$. This value agrees with numerical simulations yielding $\lambda_1(\beta_{m1})$ as $\beta_{c1}(\eta)\to 0$; see Fig.~\ref{fig7}. The mean field critical exponents are different from those of the HCVM, but they are close to them: $z=1$ is the same and the relation $\varphi=z\nu$ holds for both models \cite{gon23}.

In conclusion, we have proposed a mean field theory of the harmonically confined Vicsek model. It consists of a map for the 3D position and velocity of the swarm center of mass. The map displays transitions from quasiperiodicity to chaos and a confinement-noise phase diagram that is comparable to that of the HCVM \cite{gon23}. The phase diagram exhibits a scale-free-chaos phase transition at vanishing noise that is similar to that of the HCVM (winding number replaces polarization as order parameter). Its critical exponents are those of the Landau theory of equilibrium phase transitions \cite{hua87} plus $z=1$ and $\varphi=z\nu$ for the LLE power law. The noise shifts the chaotic interval to $(\beta_{c1}(\eta),\beta_{c2}(\eta))$ with $\lim_{\eta\to 0}\beta_{c1}(\eta)=0$ but numerical simulations show that the critical exponents are the same as in the deterministic case  \cite{suppl}. The deduced critical exponents of the mean field theory are the same for any space dimension $d>1$. We shall show elsewhere that the scale-free-chaos phase transition exists for $d=2$ but not for $d=1$. What are the consequences of the scale free phase transition for real insect swarms? This transition occurs at zero confinement and noise in the mean field theory. Its unfolding on the zero LLE noise line of Fig.~\ref{fig1}(a) informs observed behavior: qualitative shape of the swarm (on average, the center of mass rotates slowly at the rate marked by the winding number and its trajectory fills compactly a space region, akin to a condensed nucleus surrounded by vapor  \cite{sin17}; see Figures 19(b) of \cite{gon23} and S3 of \cite{suppl}), and critical exponents similar to those observed in natural swarms. The frequency of the maximal spectral amplitude and the winding number could be extracted from experimental data. As the line of zero LLE in Fig.~\ref{fig1} merges with that of the scale-free-chaos transition for infinitely many particles \cite{gon23}, it also corresponds to the same phase transition. The study of this transition at the verge of chaos for finitely many particles will be tackled in the future. A worthwhile endeavor would be reconstructing a noisy chaotic attractor from the swarm center of mass data (both in the wild or in the laboratory) by using the same techniques as in the numerical simulations of \cite{gon23}. %unfolding of the SFCPT on the zero LLE noise line informs observed behavior: compatible critical exponents, qualitative shape of the swarm (nucleus-vapor), evidences for phase transition can be reconstructed from observed power series of center of mass motion for finitely many insects.
\bigskip

{\em Acknowledgments.} This work has been supported by the FEDER/Ministerio de Ciencia, Innovaci\'on y Universidades -- Agencia Estatal de Investigaci\'on grants PID2020-112796RB-C21 (RGA) and PID2020-112796RB-C22 (LLB), by the Madrid Government (Comunidad de Madrid-Spain) under the Multiannual Agreement with UC3M in the line of Excellence of University Professors (EPUC3M23), and in the context of the V PRICIT (Regional Programme of Research and Technological Innovation). RGA acknowledges support from the Ministerio de Econom\'\i a y Competitividad of Spain through the  Formaci\'on de Doctores program Grant PRE2018-083807 cofinanced by the European Social Fund.

\bigskip
\bigskip

\centerline {\bf\large Supplementary Material}
 \renewcommand{\theequation}{S.\arabic{equation}}
\setcounter{equation}{0}
 \renewcommand{\thefigure}{S.\arabic{figure}}
\setcounter{figure}{0}

\section{Contents}\label{sec:1}
Section \ref{sec:2} contains the derivation of the mean field theory of the confined Vicsek model (VM) both for the deterministic case and for the stochastic case of nonzero alignment noise. The calculation of Lyapunov exponents in both cases is explained in Section \ref{sec:3}, which also contains a prediction of the corresponding critical exponent at the scale-free-chaos phase transition. It turns out that this phase transition occurs at zero confinement strength and zero noise. The expressions for the susceptibility and its power law near the phase transition are given in Section \ref{sec:4}. The dynamical critical exponent is obtained as a relation between the winding number and the correlation length in Section \ref{sec:5}. It turns out that the frequency at which the power spectrum reaches its maximum value coincides with the winding number. For the deterministic case, we predict the critical exponents for susceptibility, correlation length, order parameter and also the dynamical critical exponent in Section \ref{sec:6}. We find the power law for the largest Lyapunov exponent (LLE) from one lower bound we derive here and the upper bound that follows from the finite velocity of propagation of particles in the model. This critical exponent turns out to obey the same relation to dynamic and static critical exponents as found previously in the general harmonically confined VM. The numerical simulations  of Section \ref{sec:7} produce results for the critical exponents as the noise and the confinement tend to zero, as depicted in the figures of the main text. 

\section{Mean Field Vicsek model} \label{sec:2}
Consider the $d$ dimensional (dD) Vicsek model (VM)
\begin{widetext}
\begin{eqnarray}
\mathbf{x}_i(t+1)=\mathbf{x}_i(t)+ \mathbf{v}_i(t+1),\quad \mathbf{v}_i(t+1)=v_0  \mathcal{R}_\eta\!\left[\Theta\!\left(\sum_{|\mathbf{x}_j-\mathbf{x}_i|<R_0}\mathbf{v}_j(t)-\beta\mathbf{x}_i(t)\right)\!\right]\!, \label{eqa1}
\end{eqnarray}\end{widetext}
where $d=2,3$, $\Theta(\mathbf{x})=\mathbf{x}/|\mathbf{x}|$, $R_0$ is the radius of the sphere of influence about particles, $\beta$ is the confining spring constant, and $\mathcal{R}_\eta(\mathbf{w})$ performs a random rotation uniformly distributed around $\mathbf{w}$ with maximum amplitude of $\eta$ \cite{att14,gon23}. By using scale dependent Lyapunov exponents, we can determine parameter regions corresponding to deterministic chaos, noisy chaos and predominant noise \cite{gao06}, in the same way we did for the harmonically confined Vicsek model (HCVM) \cite{gon23}. 

For sufficiently small values of noise, noisy chaos has the same Lyapunov exponents as deterministic chaos \cite{gao06}. Thus, we omit the operator $ \mathcal{R}_\eta$ in Eq.~\eqref{eqa1}. Let us define the average
\begin{eqnarray}
\langle f(\mathbf{x}_i)\rangle = \frac{1}{N}\sum_{i=1}^N f(\mathbf{x}_i),\quad \mathbf{X}(t)=\langle\mathbf{x}_i\rangle,   \label{eqa2}
\end{eqnarray}
in the limit as $N\to\infty$. We now average Eq.~\eqref{eqa1} assuming the confinement takes on its critical value for the transition between single and multicluster chaos, $\beta= \beta_c(N;\eta)$ \cite{gon23}, and that the mean field approximation holds 
\begin{eqnarray}
\langle f(\mathbf{x}_i)\rangle \approx f(\langle\mathbf{x}_i\rangle).   \label{eqa3}
\end{eqnarray}
For $\eta=0$, the result is the {\em deterministic mean field HCVM}:
\begin{subequations}\label{eqa4}
\begin{eqnarray}
&&\mathbf{X}(t+1)-\mathbf{X}(t)= \mathbf{V}(t+1),\nonumber\\
&&\mathbf{V}(t+1)=v_0\Theta\!\left( \mathbf{V}(t)-\tilde{\beta}\mathbf{X}(t)\right)\!,  \quad\mbox{equivalently, } \label{eqa4a}\\
&&\mathbf{X}(t+1)-\mathbf{X}(t)\!=\! v_0\Theta\!\left( \mathbf{X}(t)-\mathbf{X}(t\!-\!1)\tilde{\beta}\mathbf{X}(t)\right)\!.\quad\quad \label{eqa4b}
\end{eqnarray}\end{subequations}
Here we have used that, for a compact swarm,
\begin{widetext}
\begin{eqnarray}
\left\langle\sum_{|\mathbf{x}_j-\mathbf{x}_i|<R_0}\mathbf{v}_j(t)\right\rangle=\left\langle\sum_{|\mathbf{x}_j-\mathbf{x}_i|<R_0}[\mathbf{x}_j(t)-\mathbf{x}_j(t-1)]\right\rangle\approx M\langle\mathbf{x}_i(t)-\mathbf{x}_i(t-1)\rangle= M [\mathbf{X}(t)-\mathbf{X}(t-1)], \label{eqa5}
\end{eqnarray}\end{widetext}
where $M$ is the average number of particles within the sphere of influence about $i$, all of which remain inside the sphere. Note that the new confining parameter is $\tilde{\beta}=\beta/M$ for Eqs.~\eqref{eqa4}. Moreover, the initial positions $\mathbf{X}(0)$ and $\mathbf{X}(1)$ characterize a plane to which all successive positions given by Eq.~\eqref{eqa4b} belong. 

We can restore the alignment noise in Eq.~\eqref{eqa4}, thereby obtaining the {\em stochastic mean field HCVM}:
\begin{widetext}
\begin{subequations}\label{eqa6}
\begin{eqnarray}
&&\mathbf{X}(t+1)=\mathbf{X}(t)+ \mathbf{V}(t+1), \quad \mathbf{V}(t+1)= v_0\mathcal{R}_\eta\!\left[ \Theta\!\left( \mathbf{V}(t)-\tilde{\beta}\mathbf{X}(t)\right)\right]\! ,\quad\mbox{equivalently, } \label{eqa6a}\\
&&\mathbf{X}(t+1)=\mathbf{X}(t)+ v_0\mathcal{R}_\eta\!\left[ \Theta\!\left( \mathbf{X}(t)-\mathbf{X}(t-1)-\tilde{\beta}\mathbf{X}(t)\right)\right]\!,\quad \tilde{\beta}=\frac{\beta}{M}. \label{eqa6b}
\end{eqnarray}
\end{subequations}\end{widetext}
Note that the polarization $W(t)=|\mathbf{V}(t)|/v_0$ always equals 1, and therefore it cannot be used as an order parameter.  We will see later that the winding number replaces it as the order parameter. 

\section{Lyapunov exponents}\label{sec:3}
Consider now a small disturbance about a trajectory of the deterministic Eq.~\eqref{eqa4b}, $\mathbf{A}\to\mathbf{A}+\delta\mathbf{A}$. We have
\begin{eqnarray}
\delta\!\left(\frac{\mathbf{A}}{|\mathbf{A}|}\right)=\left(\mathbb{I}-\frac{\mathbf{A}\mathbf{A}^T}{|\mathbf{A}|^2}\right)\!\cdot \frac{\delta\mathbf{A}}{|\mathbf{A}|}.   \label{eqa7}
\end{eqnarray}
Let $D^\pm f(t)=\mp[f(t)-f(t\pm 1)]$. Then a disturbance of Eq.~\eqref{eqa4} produces the linear equation
\begin{widetext}
\begin{subequations}\label{eqa8}
\begin{eqnarray}
\frac{D^+\delta\mathbf{X}(t)}{v_0}- \left(\mathbb{I}-\frac{\left[\left(D^--\tilde{\beta} \right)\mathbf{X}(t)\right]\left[\left(D^--\tilde{\beta}\right)\mathbf{X}(t)\right]^T}{\left|\left(D^--\tilde{\beta}\right)\mathbf{X}(t)\right|^2} \right)\cdot\frac{D^-\delta\mathbf{X}(t)-\tilde{\beta}\delta\mathbf{X}(t)}{\left|\left(D^--\tilde{\beta}\right)\mathbf{X}(t)\right|} =0 .   \label{eqa8a}
\end{eqnarray}
We now approximate $D^\pm\delta\mathbf{X}(t)\approx\delta\mathbf{\dot{X}}(t)$, thereby getting
\begin{eqnarray*}
\frac{\delta\mathbf{\dot{X}}(t)}{v_0}\!&-&\! \left(\mathbb{I}-\frac{\left[\left(D^--\tilde{\beta} \right)\!\mathbf{X}(t)\right]\left[\left(D^--\tilde{\beta}\right)\!\mathbf{X}(t)\right]^T}{\left|\left(D^--\tilde{\beta}\right)\mathbf{X}(t)\right|^2} \right)\frac{\delta\mathbf{\dot{X}}(t)}{\left|\left(D^--\tilde{\beta}\right)\mathbf{X}(t)\right|}\\
 \!&=&\! - \tilde{\beta}\!
 \left(\mathbb{I}-\frac{\left[\left(D^--\tilde{\beta} \right)\mathbf{X}(t)\right]\left[\left(D^--\tilde{\beta}\right)\mathbf{X}(t)\right]^T }{\left|\left(D^--\tilde{\beta}\right)\mathbf{X}(t)\right|^2} \right)\!\cdot \frac{\delta\mathbf{X}(t)}{\left|\left(D^--\tilde{\beta}\right)\mathbf{X}(t)\right|}, 
 \end{eqnarray*}
from which we find the linear equation
\begin{eqnarray}
&&\left(\mathbb{I}-\frac{\left[\left(D^--\tilde{\beta} \right)\mathbf{X}(t)\right]\left[\left(D^--\tilde{\beta}\right)\mathbf{X}(t)\right]^T }{\left|\left(D^--\tilde{\beta}\right)\mathbf{X}(t)\right|^2} \right)^{-1} \!\left[ \left(\mathbb{I}-\frac{\left[\left(D^--\tilde{\beta} \right)\!\mathbf{X}(t)\right]\left[\left(D^--\tilde{\beta}\right)\!\mathbf{X}(t)\right]^T}{\left|\left(D^--\tilde{\beta}\right)\mathbf{X}(t)\right|^2} \right)\right.\nonumber\\
&&-\left. \frac{\mathbb{I}}{v_0}\left|\left(D^--\tilde{\beta}\right)\!\mathbf{X}(t)\right|\right]\!\cdot  \delta\mathbf{\dot{X}}(t)= \tilde{\beta}  \delta\mathbf{X}(t)\quad\Longrightarrow \nonumber\\
&&\delta\mathbf{\dot{X}}(t) = \tilde{\beta}\!\left[ \mathbb{I}- \left(\mathbb{I}-\frac{\left[\left(D^--\tilde{\beta} \right)\!\mathbf{X}(t)\right]\left[\left(D^--\tilde{\beta}\right)\!\mathbf{X}(t)\right]^T}{\left|\left(D^--\tilde{\beta}\right)\mathbf{X}(t)\right|^2} \right)^{-1}\!\frac{\left|\left(D^--\tilde{\beta}\right)\!\mathbf{X}(t)\right|}{v_0}\right]^{-1}\!\cdot\delta\mathbf{X}(t).   \label{eqa8b}
\end{eqnarray}
Thus, we have found the matrix of Lyapunov exponents
\begin{eqnarray}
\bm{\lambda}=  \tilde{\beta}\left\langle\left[ \mathbb{I}- \left(\mathbb{I}-\frac{\left[\left(D^--\tilde{\beta} \right)\!\mathbf{X}(t)\right]\left[\left(D^--\tilde{\beta}\right)\!\mathbf{X}(t)\right]^T}{\left|\left(D^--\tilde{\beta}\right)\mathbf{X}(t)\right|^2} \right)^{-1}\!\frac{\left|\left(D^--\tilde{\beta}\right)\!\mathbf{X}(t)\right|}{v_0}\right]^{-1}\!\!\right\rangle_{t_0},   \label{eqa8c}
\end{eqnarray}
where the long time average follows whatever algorithm is used to calculate the Lyapunov exponents. From Eq.~\eqref{eqa8c}, we shall deduce later the power law 
\begin{eqnarray}
\lambda_1\sim \tilde{\beta}^{\varphi}, \label{eqa8d}
\end{eqnarray}
as $\beta\to 0$, for the largest Lyapunov exponent (LLE) with critical exponent $\varphi=0.5$. 
\end{subequations}
\end{widetext}
We now explain how to obtain the LLE by direct numerical calculation. Let us write Eq.~\eqref{eqa8a} as 
\begin{widetext}
\begin{subequations}\label{eqa9}
\begin{eqnarray}
&&\delta\bm{\chi}_{t+1}= \mathbb{M}_t\delta\bm{\chi}_t ,\quad\mathbb{M}_t=\left(\begin{array}{cc}\mathbb{I} -\tilde{\beta} \mathbb{A}_t & \mathbb{A}_t\\ - \tilde{\beta}\mathbb{A}_t &\mathbb{A}_t \\ \end{array}\right), \quad \bm{\chi}_t=\left(\begin{array}{c} \mathbf{X}(t)\\ \mathbf{V}(t) \end{array} \right)\!,  \label{eqa9a}\\
&&\mathbb{A}_t = \frac{v_0}{\left|\mathbf{V}(t)-\tilde{\beta}\mathbf{X}(t)\right|}\left[\mathbb{I}-\frac{\left(\mathbf{V}(t)-\tilde{\beta}\mathbf{X}(t)\right)\left(\mathbf{V}(t)-\tilde{\beta}\mathbf{X}(t)\right)^T}{\left|\mathbf{V}(t)-\tilde{\beta}\mathbf{X}(t)\right|^2}\right]\!. \label{eqa9b}
\end{eqnarray}
\end{subequations}
Then 
\begin{subequations}\label{eqa10}
\begin{eqnarray}
&&\frac{\delta\bm{\chi}_t}{|\delta\bm{\chi}_0|}= \mathbb{M}^t_0\frac{\delta\bm{\chi}_0}{|\delta\bm{\chi}_0|},\quad \mathbb{M}^t_0=\mathbb{M}_{t-1}\ldots\mathbb{M}_0, \quad \hat{\bm{\chi}}_t= \frac{\delta\bm{\chi}_t}{|\delta\bm{\chi}_0|}, \label{eqa10a}\\
&&h(\bm{\chi}_0,\hat{\bm{\chi}}_0) =\lim_{t\to\infty}\frac{1}{t}\ln| \mathbb{M}^t_0\hat{\bm{\chi}}_0|= \lim_{t\to\infty}\frac{1}{2t}\ln| \hat{\bm{\chi}}_0^T [\mathbb{M}^t_0]^T\mathbb{M}^t_0\hat{\bm{\chi}}_0|. \label{eqa10b}
\end{eqnarray}
\end{subequations}\end{widetext}
As the $d\times d$ matrix $\mathbb{H}^t(\bm{\chi}_0)=[\mathbb{M}^t_0]^T\mathbb{M}^t_0$ is symmetric and non-negative, its eigenvalues, i.e., the exponents $h(\bm{\chi}_0,\hat{\bm{\chi}}_0)$, are real and non-negative, and its eigenvectors are orthonormal and real. Choosing $\hat{\bm{\chi}}_0$ to be a normalized eigenvector, we obtain the corresponding eigenvalue of  $\mathbb{H}^t(\bm{\chi}_0)$, thereby the Lyapunov exponent. A random initial $\bm{\chi}_0$ produces the LLE. Similarly, eliminating a transient and selecting the corresponding vector as $\bm{\chi}_0$, Eq.~\eqref{eqa10b} yields the LLE.

A numerical approximation follows Benettin {\em et al } \cite{ben80}.  At every time $\tau_j=j\tau$, $j=0,1,\ldots,$ (the arbitrarily selected integer $\tau$ is not too large), we divide the tangent vector by its magnitude $\alpha_j$ to renormalize it to a unit length vector. Storing the $\alpha_j$, we obtain
\begin{widetext}
\begin{subequations}\label{eqa11}
\begin{eqnarray}
\lambda_1=\lim_{t\to\infty}\frac{1}{l\tau}\sum_{j=1}^l\ln\alpha_j\approx \frac{1}{l\tau}\sum_{j=1}^l\ln\alpha_j, \quad\hat{\bm{\chi}}_j= \frac{\mathbb{M}_\tau\hat{\bm{\chi}}_{j-1}}{\alpha_j}, \quad\alpha_j=|\mathbb{M}_\tau \hat{\bm{\chi}}_{j-1}|,\label{eqa11a}
\end{eqnarray}
for a sufficiently large $l$ such that the result has converged up to some tolerance. It is possible to calculate the other Lyapunov exponents by Gram-Schmidt orthogonalization \cite{ben80} or by using singular value decomposition \cite{gree87}.

We now recover the noisy rotation $\mathcal{R}_\eta$ in Eq.~\eqref{eqa1}. We obtain Eq.~\eqref{eqa10a} with $\mathcal{R}_\eta(\mathbb{M}^t_0)$ instead of $\mathbb{M}^t_0$, provided the initial condition is independent of $\eta$. In principle, Eq.~\eqref{eqa11a} should hold for every realization of $\eta$:
\begin{eqnarray}
\lambda_1=\lim_{t\to\infty}\frac{1}{l\tau}\sum_{j=1}^l\ln\alpha_j\approx \frac{1}{l\tau}\sum_{j=1}^l\ln\alpha_j, \quad\hat{\bm{\chi}}_j= \frac{\mathcal{R}_\eta(\mathbb{M}_\tau\hat{\bm{\chi}}_{j-1})}{\alpha_j}, \quad\alpha_j=|\mathcal{R}_\eta(\mathbb{M}_\tau \hat{\bm{\chi}}_{j-1})|,\label{eqa11b}
\end{eqnarray}
\end{subequations}
The equations for the stochastic mean field model are
\begin{subequations}\label{eqa12}
\begin{eqnarray}
&&\mathbf{X}(t+1)=\mathbf{X}(t)+v_0\mathcal{R}_\eta\!\left[ \Theta\!\left( \mathbf{V}(t)-\tilde{\beta}\mathbf{X}(t)\right)\right]\!, \quad\quad\label{eqa12a}\\
&&\mathbf{V}(t+1)= v_0\mathcal{R}_\eta\!\left[ \Theta\!\left( \mathbf{V}(t)-\tilde{\beta}\mathbf{X}(t)\right)\right]\!, \label{eqa12b}\\
&&\delta\bm{\chi}_{t+1}= \mathcal{R}_\eta\!\left[\mathbb{M}_t\delta\bm{\chi}_t\right]\!.  \label{eqa12c}
\end{eqnarray}
At each step, $\bm{\chi}_t$ and $\delta\bm{\chi}_t$ are known. Then a random rotation with value $\eta$ is selected. Exactly the same rotation is applied to the RHS of Eqs.~\eqref{eqa12a}, \eqref{eqa12b} and \eqref{eqa12c}. Then Eqs.~\eqref{eqa12} produce $\bm{\chi}_{t+1}$ and $\delta\bm{\chi}_{t+1}$. To include renormalization, we take $\delta\bm{\chi}_{t+1}$ given by Eq.~\eqref{eqa12c}, renormalize it
\begin{eqnarray}
\hat{\bm{\chi}}_{t+1}=\frac{\delta\bm{\chi}_{t+1}}{\alpha_{t+1}}, \quad\alpha_{t+1}=|\delta\bm{\chi}_{t+1}| ,\quad\mbox{and redefine }\,\delta\bm{\chi}_{t+1}=\hat{\bm{\chi}}_{t+1},  \label{eqa12d}
\end{eqnarray}
for the next time step. With all the values $\alpha_t$, we calculate the Lyapunov exponent as
\begin{eqnarray}
\lambda_1= \frac{1}{l}\sum_{t=1}^l\ln\alpha_t,  \label{eqa12e}
\end{eqnarray}
for sufficiently large $l$. A plot of $\lambda_1$ versus $l$ should show convergence of the exponent. Alternatively, we may eliminate a transient stage and start counting $t$ after it. See \cite{ben80} and, for the HCVM, \cite{gon23}. In the last reference, we show how to recover the same LLE from appropriate time series of the center of mass by using lagged coordinates to reconstruct the chaotic attractor.
\end{subequations}
\end{widetext}

\section{Susceptibility}\label{sec:4}
We consider the response of Eq.~\eqref{eqa6a} to an external field $\mathbf{H}$:
\begin{eqnarray}
&&\mathbf{X}(t+1)=\mathbf{X}(t)+ \mathbf{V}(t+1), \nonumber\\ 
&&\mathbf{V}(t+1)= v_0\mathcal{R}_\eta\!\left[ \Theta\!\left( \mathbf{V}(t)+\mathbf{H} - \tilde{\beta}\mathbf{X}(t)\right)\right]\!, \label{eqa13}
\end{eqnarray}
as $\mathbf{H}\to 0$. The external field is a constant force that would appear in the equations of motion if we add a potential $-\mathbf{H}\cdot\sum_{j=1}^N\mathbf{x}_j(t)$ to a Hamiltonian of the system of particles. The function equivalent to magnetization in the Ising model is the position of the center of mass given by Eq.~\eqref{eqa2}. Its average magnitude is $\langle R(t)\rangle_t$. We define the response matrix $\mathbf{\mathcal{H}}_t=\nabla_\mathbf{H}\bm{\chi}_t= \partial\bm{\chi}_t^i/\partial H_j$ (at zero field). Then the first equation in Eq.~\eqref{eqa13} produces
\begin{widetext}
\begin{subequations}\label{eqa14}
\begin{eqnarray} 
\mathbb{Y}_{t+1}= \mathbb{Y}_t+ \mathbb{W}_{t+1},\quad \mbox{ where }\quad\mathbf{\mathcal{H}}^{ij}=\left.\!\left(\begin{array}{c}\frac{\partial\mathbf{X}_i}{\partial H_j}\\ \frac{\partial\mathbf{V}_i}{\partial H_j}\end{array}\right)\right|_{\mathbf{H}=\mathbf{0}} =  \left(\begin{array}{c} \mathbb{Y}\\ \mathbb{W} \end{array}\right)\!, \quad (\mathbf{Y}^j )_i=\mathbb{Y}^{ij}, \quad (\mathbf{W}^j )_i=\mathbb{W}^{ij},  \label{eqa14a}
\end{eqnarray}
and the second equation in Eq.~\eqref{eqa13},
\begin{eqnarray} 
(\mathbf{W}_{t+1}^j)_i=\mathcal{R}_\eta\!\left(\mathbb{A}_t^{ik}\!\left[(\mathbf{W}^j_t)_k -\tilde{\beta}(\mathbf{Y}_t^j)_k\right]\! + \mathbb{A}_t^{ik}(\bm{\delta}^j)_k\right)\!,\quad (\bm{\delta}^j)_i= \delta_{ij}.  \label{eqa14b}
\end{eqnarray}
Here sum over repeated indices is understood. The size of the response matrix at zero field is 
\begin{eqnarray}
\chi=\langle\lVert\mathbf{\mathcal{H}}_t\rVert\rangle_t, \quad \lVert\mathbf{\mathcal{H}}_t\rVert =\sqrt{\lambda_M(\mathbf{\mathcal{H}}_t\mathbf{\mathcal{H}}_t^T)}, \label{eqa14c}
\end{eqnarray}\end{subequations}\end{widetext}
where $\lambda_M(\mathbf{\mathcal{H}}_t\mathbf{\mathcal{H}}_t^T)$ is the maximum eigenvalue of the symmetric positive matrix $\mathbf{\mathcal{H}}_t\mathbf{\mathcal{H}}_t^T$ and $\langle\ldots\rangle_t$ is a time average. We find the same results replacing $\mathbb{Y}_t$ instead of $\mathbf{\mathcal{H}}_t$ in Eq.~\eqref{eqa14c}. In the limit of vanishing field, $\chi$ in Eq.~\eqref{eqa14c} can be estimated by $\chi\sim \partial\langle R\rangle_t/\partial H$ as $H\to 0$, where $\langle R\rangle_t$ plays a role analogous to average magnetization in the Ising model \cite{ami05}. Thus, we find the scalar susceptibility $\chi$ by numerical evaluation of Eq~\eqref{eqa14c} and obtain a quantitative measure of the usual scalar susceptibility. We shall find an estimation thereof in Section \ref{sec:6}. The scale-free-chaos phase transition occurs as $\beta\to 0+$. Thus, we have the power laws 
\begin{eqnarray}
\langle R(t)\rangle_t\sim \tilde{\beta}^{-\nu},  \quad \chi\sim \tilde{\beta}^{-\gamma},\label{eqa15}
\end{eqnarray}
where $R(t)=|\mathbf{X}(t)|$, and $ \langle R(t)\rangle_t$ is the time averaged radius of the center of mass, which plays the role of correlation length. Alternatively, we could select $L=$max$_tR(t)$ as the correlation length because $L$ is proportional to $\langle R(t)\rangle_t$ as shown by Fig.~5(a) of the main text.

\section{Power spectrum}\label{sec:5}
Given a signal $s(t)=X(t)+Y(t)+Z(t)$, $t=1,\ldots,N$, the discrete Fourier transform is
\begin{eqnarray}
&&\hat{s}(\omega)=\sum_{t=1}^N s(t)\, e^{i2\pi t\omega/N},\quad s(t)=\frac{1}{N}\sum_{\omega=1}^N\hat{s}(\omega)\, e^{-i2\pi \omega t/N},\nonumber\\
&&s(t\pm N)=s(t),\, \hat{s}(\omega\pm N)=\hat{s}(\omega). \label{eqa16}
\end{eqnarray}
Thus, $s(0)=s(N)$, $\hat{s}(0)=\hat{s}(N)$. For a real valued signal, $\hat{s}(\omega)=\overline{\hat{s}(N-\omega)}$, and we have the autocorrelation function $C(\omega)$:
\begin{eqnarray}
&&C(\tau)=\sum_{t=1}^N s(t)s(t+\tau)= \frac{1}{N}\sum_{\omega=1}^N |\hat{s}(\omega)|^2 e^{-i2\pi \omega\tau/N},\nonumber\\
&& |\hat{s}(\omega)|^2= \sum_{\tau=1}^N C(\tau)\, e^{i2\pi \omega\tau/N}. \label{eqa17}
\end{eqnarray}
The graph of $ |\hat{s}(\omega)|^2$ as a function of the frequency $\nu(\omega)=\omega/N$ is the power spectrum \cite{ras90}. The power spectrum of a chaotic signal typically exhibits noisy behavior at low frequencies. All the power is expected to be in the low frequencies because aperiodic points in a finite data set appear as points with very long periods, comparable to the total sample time, and consequently, correspond to very low frequencies \cite{ras90}.

In terms of the highest peak of the power spectrum at frequency $\Omega=$ argmax$_\omega |\hat{s}(\nu(\omega))|^2$, we have the power law
\begin{eqnarray}
\Omega\sim \langle R(t)\rangle_t^{-z}, \label{eqa18}
\end{eqnarray}
at the critical confinement as $\beta\to 0$. Here $z$ is the dynamical critical exponent. For $\eta=0$ and within the first chaotic window for $\eta>0$, numerical evidence shows that the frequency $\Omega$ coincides with the winding number defined as
\begin{eqnarray}
w=\lim_{n\to\infty} \frac{1}{2\pi n}\sum_{k=1}^n\theta_k, \label{eqa19}
\end{eqnarray}
where $\theta_k$ is the angle between vectors $\mathbf{X}(k)$ and $\mathbf{X}(k+1)$. See Figure 4 of the main text.

\section{Mean field critical exponents}\label{sec:6}
From Eq.~\eqref{eqa4a}, we obtain
\begin{widetext}
\begin{eqnarray*}
v_0^2=|\mathbf{X}(t)|^2 +|\mathbf{X}(t+1)|^2 - 2|\mathbf{X}(t)|\,|\mathbf{X}(t+1)|\cos\theta(t+1)=R(t)^2+R(t+1)^2-2R(t)R(t+1)\cos\theta(t+1).
\end{eqnarray*}\end{widetext}
We now average over time and ignore fluctuations. Then
\begin{eqnarray*}
&&\langle R(t)^2\rangle_t\approx\langle R(t)\rangle_t^2, \\
&&\langle R(t)R(t+1)\cos\theta(t+1)\rangle_t\approx \langle R(t)\rangle_t^2\cos\langle\theta(t)\rangle_t.
\end{eqnarray*}
According to Eq.~\eqref{eqa19}, the time-averaged angle $\theta(t)$ is $2\pi w$. Thus, we get
\begin{eqnarray}
v_0^2= 2\langle R\rangle_t^2[1-\cos(2\pi w)]. \label{eqa20}
\end{eqnarray}
As we approach the critical confinement, $\beta\to 0+$, $\langle R(t)\rangle_t\to\infty$ and therefore $w\to 0$ so that  
\begin{eqnarray}
w\sim \frac{v_0}{2\pi \langle R\rangle_t}. \label{eqa21}
\end{eqnarray}
Thus, the dynamical critical exponent of Eq.~\eqref{eqa18} with $\Omega=w$ is $z=1$. Multiplying Eq.~\eqref{eqa4b} by $\mathbf{X}(t)$, we find:
\begin{widetext}
\begin{eqnarray*}
R(t+1)R(t)\cos\theta(t+1)- R(t)^2=v_0\frac{(1-\tilde{\beta})R(t)^2-R(t)R(t-1)\cos\theta(t)}{\sqrt{(1-\tilde{\beta})^2R(t)^2+R(t-1)^2 -2R(t)R(t-1) (1-\tilde{\beta})\cos\theta(t)}}.
\end{eqnarray*}
Time averaging this expression and ignoring fluctuations, we find
\begin{eqnarray*}
&&-[1-\cos(2\pi w)]\langle R\rangle_t^2= v_0\langle R\rangle_t^2\frac{1-\tilde{\beta} -\cos(2\pi w)}{\sqrt{2(1-\tilde{\beta})\langle R\rangle_t^2[1-\cos(2\pi w)]+\tilde{\beta}^2\langle R\rangle_t^2}}\Longrightarrow\\
&&\sqrt{1-\tilde{\beta}+\frac{\tilde{\beta}^2\langle R\rangle_t^2}{v_0^2}}=\frac{\tilde{\beta}}{1-\cos(2\pi w)}-1= \frac{2\tilde{\beta}\langle R\rangle_t^2}{v_0^2}-1,
\end{eqnarray*}
where we have used Eq.~\eqref{eqa20}. After some algebra, the square of this expression produces
\begin{eqnarray*}
\tilde{\beta}\left(\frac{4\langle R\rangle_t^2}{v_0^2}-1\right)\left(1-\frac{\tilde{\beta}\langle R\rangle_t^2}{v_0^2}\right)=0.
\end{eqnarray*}
Then we obtain
\begin{eqnarray}\label{eqa22}
\langle R\rangle_t= \frac{v_0}{\sqrt{\tilde{\beta}}},\quad  w\sim\frac{\sqrt{\tilde{\beta}}}{2\pi},
\end{eqnarray}
in which we have used Eq.~\eqref{eqa21} in the limit as $\tilde{\beta}\to 0+$. Thus, we have found the relation $w\sim\tilde{\beta}^b$, with $b=1/2$, for the winding number, which plays the role of order parameter in the Landau theory. Eq.~\eqref{eqa22} also produces the critical exponent $\nu=1/2$ in Eq.~\eqref{eqa15}.  Notice that the previous relations also produce
\begin{eqnarray}
\langle|\mathbf{V}(t)-\tilde{\beta}\mathbf{X}(t)|\rangle_t=[1 + o(\tilde{\beta})]v_0. \label{eqa23}
\end{eqnarray}

To get the exponent for the susceptibility, we note that the size of the response matrix \eqref{eqa14c} is given by the derivative of $\langle R\rangle_t=\langle|\mathbf{X}(t+1)|\rangle_t$. From Eq.~\eqref{eqa13} for $\eta=0$,
 \begin{eqnarray*}
R(t+1)^2=R(t)^2+v_0^2+2v_0\frac{(1-\tilde{\beta})R(t)^2-R(t)R(t-1)\cos\theta(t)+HR(t)\cos\Theta(t)}{\sqrt{(1-\tilde{\beta})^2R(t)^2+R(t-1)^2 -2R(t)R(t-1) (1-\tilde{\beta})\cos\theta(t)+2Hv_0\cos\phi(t)+H^2}}, 
\end{eqnarray*}
Time averaging, ignoring fluctuations, and setting $\langle\Theta\rangle_t= \langle\phi\rangle_t=0$, we find 
\begin{eqnarray*}
-v_0^2=2v_0 \frac{[1-\tilde{\beta} -\cos(2\pi w)] \langle R\rangle_t^2+H\langle R\rangle_t}{\sqrt{2\langle R\rangle_t^2(1-\tilde{\beta})[1-\cos(2\pi w)]+\tilde{\beta}^2\langle R\rangle_t^2-2\tilde{\beta}H+H^2}}= \frac{v_0^2 -2\tilde{\beta}\langle R\rangle_t^2+2H\langle R\rangle_t}{\sqrt{1-\tilde{\beta} +\frac{\tilde{\beta}^2\langle R\rangle_t^2}{v_0^2}- \frac{2\tilde{\beta}H}{v_0^2}+\frac{H^2}{v_0^2}}}.
\end{eqnarray*}
We now differentiate this expression with respect to $H$ and set $H=0$. The result is
\begin{eqnarray*}
0=2\langle R\rangle_t-4\tilde{\beta}\langle R\rangle_t\frac{\partial\langle R\rangle_t}{\partial H} - (v_0^2-2\tilde{\beta}\langle R\rangle_t^2)\left(\frac{\tilde{\beta}^2\langle R\rangle_t }{v_0^2}\frac{\partial\langle R\rangle_t}{\partial H}- \frac{2\tilde{\beta}}{v_0^2} \right)\!\sim \frac{2v_0}{\sqrt{\tilde{\beta}}}- 4v_0\tilde{\beta}^{1/2}\frac{\partial\langle R\rangle_t}{\partial H}+v_0\tilde{\beta}^{3/2}\frac{\partial\langle R\rangle_t}{\partial H}-2\tilde{\beta},
\end{eqnarray*}\end{widetext}
where we have used Eq.~\eqref{eqa22}. Dominant balance produces
\begin{eqnarray}
\left.\frac{\partial\langle R\rangle_t}{\partial H}\right|_{H=0}\sim \frac{1}{2\tilde{\beta}}. \label{eqa24}
\end{eqnarray}
This equation gives the order of magnitude of the susceptibility defined by Eq.~\eqref{eqa14c} as the size of the response matrix at zero external field. The critical exponent for the susceptibility is therefore $\gamma=1$. Thus, the critical exponents of the deterministic mean field model of Eqs.~\eqref{eqa4} are the same as those in the Landau theory of phase transitions:
\begin{equation}
\nu=b=\frac{1}{2}, \quad z=\gamma=1. \label{eqa25}
\end{equation}

To find the critical exponent $\varphi$, we first use the triangular inequality in Eq.~\eqref{eqa4a} and then Eq.~\eqref{eqa22},
\begin{widetext}
\begin{eqnarray*}
&&v_0=\langle|\mathbf{V}(t+1)|\rangle_t=\langle|\mathbf{V}(t+1)-\tilde{\beta}\mathbf{X}(t)+\tilde{\beta}\mathbf{X}(t)|\rangle_t\leq \tilde{\beta}\langle R\rangle_t+ \langle|\mathbf{V}(t+1)-\tilde{\beta}\mathbf{X}(t)|\rangle_t\Longrightarrow \\
&&\langle|\mathbf{V}(t+1)-\tilde{\beta}\mathbf{X}(t)|\rangle_t\geq v_0(1-\sqrt{\tilde{\beta}})\Longrightarrow 1- \frac{\langle|\mathbf{V}(t+1)-\tilde{\beta}\mathbf{X}(t)|\rangle_t}{v_0}\leq \sqrt{\tilde{\beta}}.
\end{eqnarray*}\end{widetext}
Time-averaging Eq.~\eqref{eqa8c} with $\langle\mathbf{X}(t)\rangle_t=0$ and using the entry for the LLE, we get
\begin{eqnarray}
\lambda_1\geq \sqrt{\tilde{\beta}}\Longrightarrow \varphi\leq 0.5, \label{eqa26}
\end{eqnarray}
provided $\lambda_1\sim\tilde{\beta}^\varphi$.  Together with the inequality $\varphi\geq\nu$ deduced in \cite{gon23} and the critical exponents of Eq.~\eqref{eqa25}, this shows that $\varphi=0.5$. 

\section{Numerical results}\label{sec:7}
Figure 1 of the main text shows the regions of the confinement vs noise plane corresponding to different attractors. For very small $\eta$, and increasing $\beta$, deterministic chaos changes to quasiperiodic attractors, and then to period 4 and period 2 attractors. For larger noise values, the previously enumerated attractors become modified by noise and another window of chaos may appear. For $2<\eta<2\pi$, noise dominates even though there is a region of chaos swamped by noise for intermediate values of $\beta$. For the mean field VM, the scale-free-chaos phase transition of the confined VM corresponds to the origin $(0,0)$ of the phase plane in Figure 1 of the main text. This point is the limit as $\beta\to 0+$ of the deterministic model with $\eta=0$. The predicted critical exponents are those of Eq.~\eqref{eqa25}. 

\begin{figure}[h]
\begin{center}
\includegraphics[width=7cm]{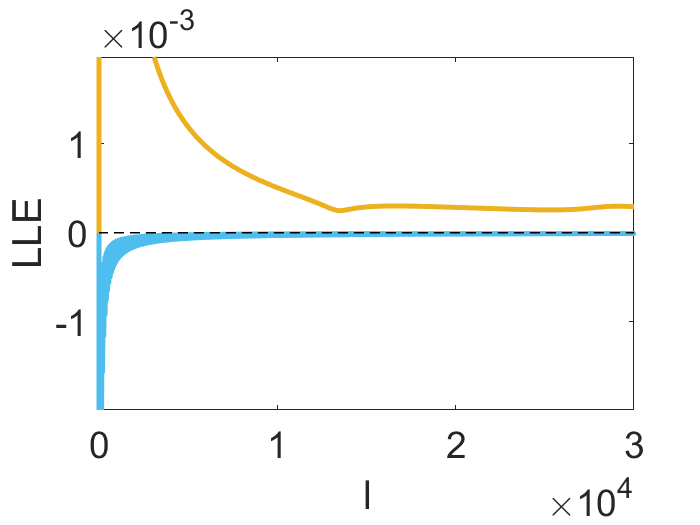}
\end{center}
\caption{Largest Lyapunov exponents as functions of $l$ as given by Eq.~\eqref{eqa12e} for $\eta=0$ and $\beta=10^{-7}$ ($\lambda_1>0$, orange line) and $\beta=1$ ($\lambda_1\leq 0$, blue line). \label{figS1}}
\end{figure}

Figure 2 of the main text shows the windows of positive largest Lyapunov exponent (LLE) for vertical lines in Fig. 1 of the main text at noises $\eta=0$ and $\eta=0.5$, corresponding to deterministic and noisy chaos, respectively. For zero noise, the chaotic window begins at $\beta_{c1}=0$ (we have been able to reach down to $\beta=10^{-9}$ and still get a clearly positive LLE within the range of our numerical simulations, see Fig.~\ref{figS1}) and ends at $\beta_{c2}=0.2$. Fig.~2(b) of the main text indicates that there are two chaotic windows $(\beta_{c1},\beta_{c2})$ and $(\beta_{c3},\beta_{c4})$ for $\eta=0.5$. Inside these chaotic windows, the maximum LLE is reached at single values $\beta_{m1}$ and $\beta_{m2}$, respectively. Fig. 1 of the main text shows that the scale-free-chaos phase transition is located at the origin of the phase diagram $(\beta,\eta)$. While we can reach this transition by lowering $\beta$ at $\eta=0$, we can also move on the critical line $\beta_{c1}(\eta)$ in Fig. 1 of the main text and let $\eta\to 0$ until we end at the origin of the phase diagram. This latter route to the scale-free-chaos phase transition is reminiscent of finite size scaling for the confined VM, in which we find critical exponents by letting $N\to\infty$ on the critical lines $\beta_c(N;\eta)$ as $\beta_c\to 0+$ for fixed $\eta$ on the region of noisy chaos \cite{gon23}. The critical exponents obtained by either route are the same but the deterministic route allows us to build the theory explained in previous sections, whereas the critical exponents obtained by descending through the noise line $\beta_{c1}(\eta)$ in Fig. 1 of the main text follow from numerical simulations.

\begin{widetext}
Figure 3(a) of the main text shows different chaotic attractors for increasing values of $\beta$ (with $v_0=M=1$, which is equivalent to replacing $\tilde{\beta}$ instead of $\beta$ in the figures). As $\tilde{\beta}$ increases past $\tilde{\beta}_{m1}(0)$ (corresponding to the maximum of the LLE), the attractors shrink to an annulus that fills a fraction of the space they did for smaller values of  $\tilde{\beta}$. The shape of these chaotic attractors lends support to the hypothesis $\langle\mathbf{X}(t)\rangle_t=\langle\mathbf{V}(t)\rangle_t= 0$ made when simplifying Eq.~\eqref{eqa8c} by ignoring fluctuations in the mean field theory. 
\begin{figure}[h]
\begin{center}
\includegraphics[width=4.4cm]{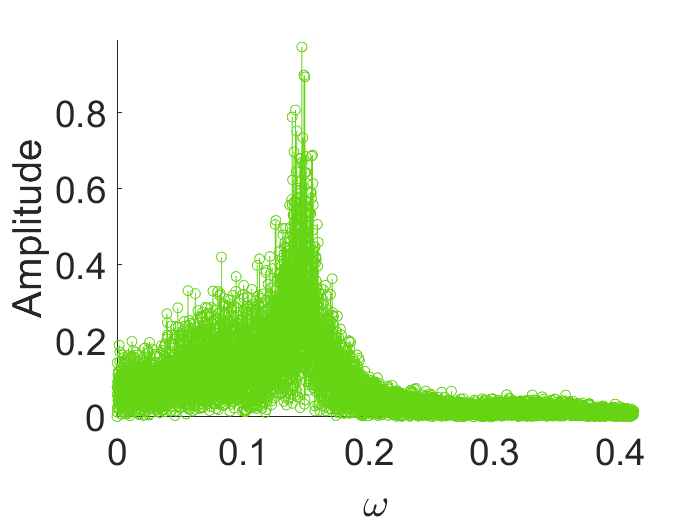}
\includegraphics[width=4.4cm]{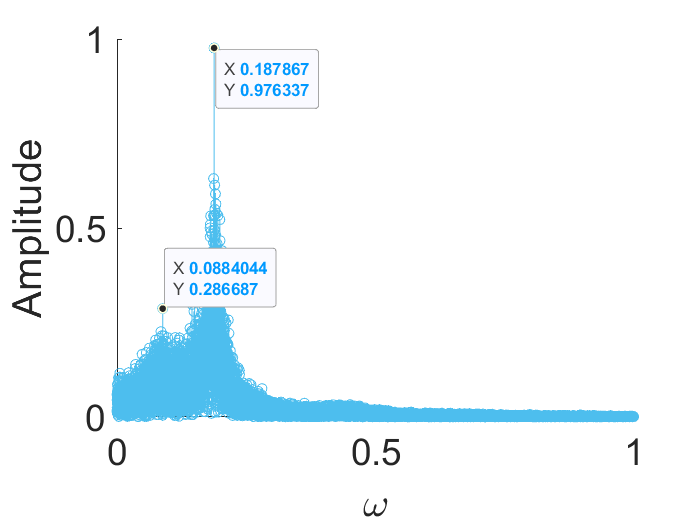}
\includegraphics[width=4.4cm]{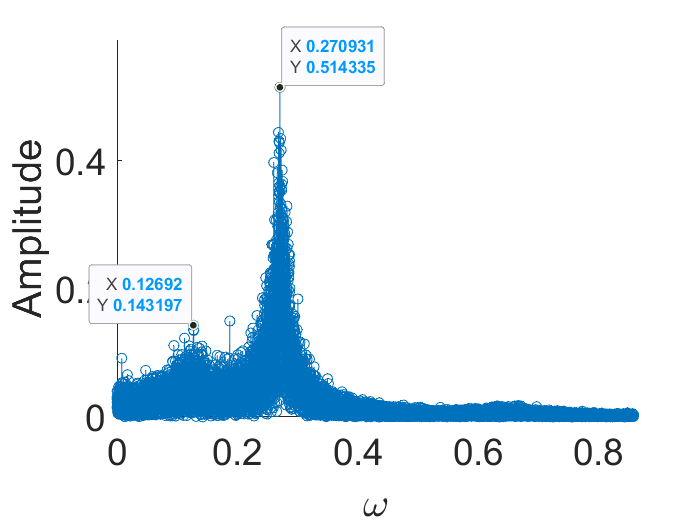}
\includegraphics[width=4.4cm]{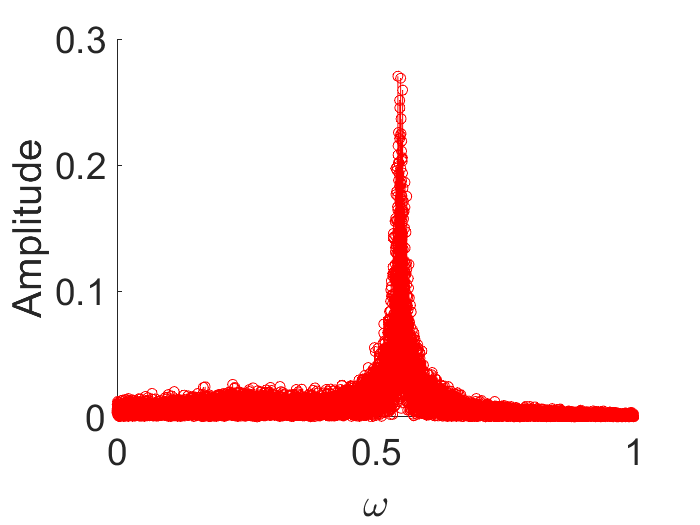}\\
{\bf (a)} \hskip 4cm {\bf (b)} \hskip 4cm {\bf (c)}\hskip 4.5cm {\bf (d)}
\end{center}
\caption{Amplitude of the power spectrum vs frequency for $\eta=0.5$ and $\beta$ values belonging to the first chaotic window in Figures 2(b) and 4(b) of the main text: {\bf (a)} $\beta=0.025$, {\bf (b)} $\beta=\beta_{m1}(0.5)=0.04$, {\bf (c)} $\beta=0.08$, and {\bf (d)}  $\beta=0.3$. For these values, the frequency of the maximum amplitude coincides with the winding number. In panels (b) and (c) there is an appreciable secondary maximum, which does not surpass the main one.} \label{figS2}
\end{figure}
\end{widetext}

Figure 4(a) of the main text depicts the winding number of Eq.~\eqref{eqa19} for $\eta=0$, which coincides with the frequency $\Omega$ corresponding to the largest peak in the Fourier spectrum of the signal $s=X+Y+Z$. In the presence of noise, the situation is more complex as shown in Fig. 4(b) of the main text. Within the first chaotic window, $(\beta_{c1},\beta_{c2})$, we have $\Omega=w$, whereas $\Omega<w$ for $\beta<\beta_{c1}$. Figures \ref{figS2} depict the amplitudes of the power spectrum versus frequency for different values of $\beta$ within the first chaotic window at $\eta=0.5$. The power spectrum exhibits the noisy behavior at low frequencies that is typical of chaos \cite{ras90}. While a secondary amplitude maximum appears at a smaller frequency for intermediate values of $\beta$, it does not surpass the main one. Within the second chaotic window, $(\beta_{c3},\beta_{c4})$, $\Omega$ is piecewise constant, with a finite jump at $\beta_{m2}$, which corresponds to the maximum of the LLE. The winding number is smooth: it is slightly larger than $\Omega$ for  $\beta_{c3}<\eta<\beta_{m2})$, and it is slightly smaller than $\Omega$ for  $\beta_{m2}<\beta< \beta_{c4})$; see the inset of Fig. 4(b) of the main text.

Figure 5 of the main text depicts different power laws for the scale-free-chaos phase transition. For the deterministic case $\eta=0$, Fig.~5(a) of the main text shows that the swarm size  $L=\max_tR(t)$ is proportional to the correlation length, which is the time averaged length of the center of mass position as the confinement decreases. This indicates a scale free transition. Fig.~5(b) of the main text plots the power law of the correlation length $\xi=\langle R\rangle_t$ versus $\beta$, $\xi\sim \beta^{-\nu}$, with $\nu=0.5$. The dynamical critical exponent $z$ is found from the relation $\Omega\sim\xi^{-z}$, $z=1.01\pm 0.01$, between the frequency corresponding to the maximum of the power spectrum and the correlation length; see Fig.~5(c) of the main text. Lastly the power law for the susceptibility, $\chi\sim\beta^{-\gamma}$, with $\gamma=1.000\pm 0.002$, is shown in Fig.~5(d) of the main text.

Figure 6 of the main text depicts the power laws for the stochastic case as we approach the origin of the phase diagram (Figure 1 of the main text) through the line $\beta_{c1}(\eta)$ that marks the beginning of the first chaotic window. We check that the critical exponents of Eq.~\eqref{eqa25} are also obtained as $\beta\to 0$ on the line $\beta_{c1}(\eta)$. Since this line marks the beginning of the first chaotic window, the LLE is zero on it, and we cannot determine the critical exponent $\varphi$ for nonzero noise from it.

To determine the critical exponent $\varphi$, we plot the maximum value of the LLE $\lambda_1$ for each value of the noise used to draw the line $\beta_{c1}(\eta)$. The result is Figure 7 of the main text and we find $\varphi=0.45\pm 0.05$, which is compatible with the predicted exponent $\varphi=0.5$ of Eq.~\eqref{eqa26}.

The critical exponents obtained from numerical simulations of the HCVM are $\nu=0.436$, $\gamma= 0.92$, $z=1$, $b=0.58$, $\varphi\approx z\nu$ \cite{gon23}. As expected, the values of several critical exponents are different from (but relatively close to) those of the MFHCVM. However, since $z=1$ and $\nu=0.5$, the relation $\varphi=z\nu$, deduced for the HCVM \cite{gon23}, also holds for the mean field model within error bounds. Measured critical exponents are $\nu=0.35$, $\gamma=0.9$, \cite{att14} $z=1.12$ \cite{cav17}; see also \cite{cav21arxiv}.
\begin{figure}[h]
\begin{center}
\includegraphics[width=7cm]{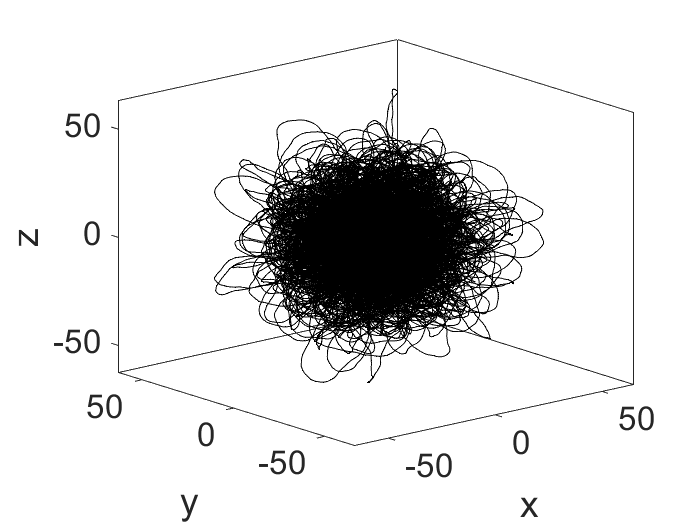}
\end{center}
\caption{The trajectory of the center of mass at the verge of chaos ($\eta=0.5$ and $\beta=\beta_{c1}=0.002$) fills a compact 3D region. \label{figS3}}
\end{figure}

For the deterministic case, the center of mass moves on a plane, as illustrated by Figure 3(a) of the main text. However, for appreciable noise, $\eta=0.5$, the center of mass trajectory fills a compact volume whose boundary is less visited, as shown in Fig.~\ref{figS3}. This figure, obtained solving the mean field model at the critical line $\beta_{c1}(\eta)$ of zero LLE, is similar to Figure 19(b) of \cite{gon23} for the HCVM in the noisy chaos region.

\end{document}